\documentclass[preprint2]{aastex}

\usepackage{graphicx}
\usepackage{natbib}
\usepackage[figuresleft]{rotating}
\newcommand{\HI}{H{\,\small I}}

\newcommand{\Htwo}{H$_{2}$}
\newcommand{\ltsima} {$\; \buildrel < \over \sim \;$}
\newcommand{\gtsima} {$\; \buildrel > \over \sim \;$}
\newcommand{\lta} {\lower.5ex\hbox{\ltsima}}
\newcommand{\gta} {\lower.5ex\hbox{\gtsima}}

\newcommand{\kms}{km\,s$^{-1}$}

\newcommand{\lya}{Ly$\alpha$}

\shorttitle{CO(1-0) in $z \sim 2$ radio galaxy MRC~0152-209}
\shortauthors{Emonts et al.}

\begin{document}


\title{Molecular CO(1-0) gas in the $z \sim 2$ radio galaxy MRC~0152-209}


\author{B. H. C. Emonts\altaffilmark{1,12}, I. Feain\altaffilmark{1}, M. Y. Mao\altaffilmark{2,1,3}, R. P. Norris\altaffilmark{1}, G. Miley\altaffilmark{4}, R. D. Ekers\altaffilmark{1}, M. Villar-Mart\'{i}n\altaffilmark{5}, H. J. A. R\"{o}ttgering\altaffilmark{4}, E. M. Sadler\altaffilmark{6}, G. Rees\altaffilmark{7,1}, R. Morganti\altaffilmark{8,9}, D. J. Saikia\altaffilmark{10}, T. A. Oosterloo\altaffilmark{8,9}, J. B. Stevens\altaffilmark{1}, C. N. Tadhunter\altaffilmark{11}}

\affil{$^{1}$CSIRO Astronomy and Space Science, Australia Telescope National Facility, PO Box 76, Epping NSW, 1710, Australia} 
\affil{$^{2}$School of Mathematics and Physics, University of Tasmania, Private Bag 37, Hobart, 7001, Australia}
\affil{$^{3}$Australian Astronomical Observatory, PO Box 296, Epping, NSW, 1710, Australia}
\affil{$^{4}$Leiden Observatory, University of Leiden, P.O. Box 9513, 2300 RA Leiden, Netherlands}
\affil{$^{5}$Instituto de Astrof\'{i}sica de Andaluc\'{i}a (CSIC), Aptdo. 3004, Granada, Spain}
\affil{$^{6}$School of Physics, University of Sydney, NSW 2006, Australia}
\affil{$^{7}$School of Physics, Astronomy and Mathematics, University of Hertfordshire, College Lane, Hatfield AL10 9AB}
\affil{$^{8}$Netherlands Institute for Radio Astronomy, Postbus 2, 7990 AA Dwingeloo, the Netherlands}
\affil{$^{9}$Kapteyn Astronomical Institute, University of Groningen, P.O. Box 800, 9700 AV Groningen, the Netherlands}
\affil{$^{10}$National Centre for Radio Astrophysics, TIFR, Pune University Campus, Post Bag3, Pune 411 007, India}
\affil{$^{11}$Department of Physics and Astronomy, University of Sheffield, Sheffield S3 7RH, UK}
\affil{$^{12}$Bolton Fellow}
\email{bjorn.emonts@csiro.au}




\begin{abstract}
We report the detection of molecular CO(1-0) gas in the high-$z$ radio galaxy MRC~0152-209 ($z = 1.92$) with the Australia Telescope Compact Array Broadband Backend (ATCA/CABB). This is the third known detection of CO(1-0) in a high-$z$ radio galaxy to date. CO(1-0) is the most robust tracer of the overall molecular gas content (including the wide-spread, low-density and subthermally excited component), hence observations of CO(1-0) are crucial for studying galaxy evolution in the Early Universe. We derive $L'_{\rm CO} = 6.6 \pm 2.0 \times 10^{10}$ ${\rm K~km~s^{-1}~pc^2}$ for MRC~0152-209, which is comparable to that derived from CO(1-0) observations of  several high-$z$ submillimeter and starforming $BzK$ galaxies. The CO(1-0) traces a total molecular hydrogen mass of M$_{\rm H2} = 5 \times 10^{10}$ ($\alpha_{\rm x}$/0.8) M$_{\odot}$. MRC~0152-209 is an infra-red bright radio galaxy, in which a large reservoir of cold molecular gas has not (yet) been depleted by star formation or radio source feedback. Its compact radio source is reliably detected at 40~GHz and has a steep spectral index of $\alpha = -1.3$ between 1.4 and 40 GHz ($4-115$ GHz in the galaxy's rest-frame). MRC~0152-209 is part of an ongoing systematic ATCA/CABB survey of CO(1-0) in high-$z$ radio galaxies between $1.7 < z < 3$.

\end{abstract}





\section{Introduction}
\label{sec:introduction}

High-redshift radio galaxies (HzRGs; $L_{\rm 500 MHz} > 10^{27}$ W~Hz$^{-1}$) are among the most massive galaxies in the early Universe \citep[][]{mil08}. They are typically surrounded by rich proto-cluster environments and are believed to be the ancestors of early-type central-cluster galaxies \citep[e.g.][]{ven07}. HzRGs are in an active stage of their evolution, with powerful radio jets emanating from a massive central black-hole. These radio jets vigorously interact with their surrounding environment \citep[e.g.][]{hum06} and serve as beacons for tracing the faint host galaxy and surrounding proto-cluster. This makes HzRGs among the best studied high-$z$ objects and ideal laboratories for investigating both the formation and evolution of galaxies/clusters as well as the relationship between early star formation and AGN activity.

A crucial component in the evolutionary studies of HzRGs and their associated star formation processes is a detailed knowledge about the content and properties of the molecular gas. The most abundant molecule, molecular hydrogen or \Htwo, has strongly forbidden rotational transitions and is virtually invisible, except when it is shocked or heated to high temperatures. An excellent tracer for \Htwo\ gas is carbon monoxide or CO, which emits strong rotational transition lines that occur primarily through collisions with \Htwo, even at relatively low densities \citep[e.g.][]{sol05}. \citet{bro91} were the first to observe $^{12}$CO (referred to as CO in this paper) beyond $z=2$. 

During that same decade, intensive searches failed to detect CO in HzRGs \citep{eva96,oji97}. Since then, CO emission has been detected in individual studies of powerful radio galaxies between $z \sim 2-5$ \citep[see reviews by][and Sect. \ref{sec:discussion2} for individual references]{sol05,mil08}. In some cases the CO has been resolved on scales of several tens of kpc \citep[e.g.][]{pap00}. 

Despite these results, two major limitations to systematic searches for CO in HzRGs have been the very limited velocity coverage of existing mm-spectrometers (often not much wider than the velocity range of the CO gas and/or the accuracy of the redshift) and the fact that most observatories can only target the higher rotational transitions of CO at high-$z$. While these higher CO transitions trace the dense and thermally excited gas (such as that in the nuclear starburst/AGN region), \citet[][]{pap00,pap01} suggest that more widely distributed reservoirs of less dense and sub-thermally excited gas could be largely missed by these observations and could thus more easily be detected in the lower transitions than often assumed \citep[see also e.g.][and references therein]{dan09,car10,ivi11}. The luminosity of the ground transition, CO(1-0), is least affected by the excitation conditions of the gas. Therefore, observations of CO(1-0) are crucial for deriving the most accurate estimates of the overall molecular gas content in high-$z$ galaxies.

Ongoing developments in broadband correlators are overcoming the above mentioned limitations. As we discussed in \citet{emo11}, the upgrade of the Australia Telescope Compact Array (ATCA) with the $2 \times 2$\,GHz Compact Array Broadband Backend \citep[CABB;][]{wil11} now allows searching for CO(1-0) across the southern hemisphere. We are currently in the process of observing an unbiased sample of HzRGs from the Molonglo Reference Catalogue \citep{mcc90} in CO(1-0) with the ATCA/CABB 7mm system between $1.7<z<3$.

In this Letter, we present the detection of CO(1-0) emission in one of our sample sources, MRC\,0152-209 ($z$=$1.92$). MRC\,0152-209 is a radio galaxy with a disturbed optical morphology \citep{pen01}. It is bright in the near- and mid-IR \citep{sey07,bre10}, detected in \lya\ \citep{mcc91} and has a compact radio source \citep[1.6~arcsec/13~kpc in diameter;][]{pen00_radio}. 

Throughout this paper, we assume H$_{0} = 71$\,\kms\,Mpc$^{-1}$, $\Omega_{\rm M} = 0.3$ and $\Omega_{\Lambda} = 0.7$, corresponding to an angular distance scale of 8.3 kpc arcsec$^{-1}$ and luminosity distance $D_{\rm L} = 14571$ Mpc for MRC~0152-209 \citep[following][]{wri06}.\footnote{See http://www.astro.ucla.edu/$\sim$wright/CosmoCalc.html}

\section{Observations}
\label{sec:observations}

MRC~0152-209 was observed on 25 $\&$ 26 August and 28 $\&$ 29 September 2010 with ATCA/CABB in the most compact hybrid H168 and H75 array configurations. The total effective on-source integration time was 15.5h (with 45$\%$ of our overall observing time spent on overheads and calibration). We used the coarsest spectral-line CABB mode ($2 \times 2$\,GHz bandwidth; 1\,MHz resolution) and centered both 2 GHz observing bands around 39.476 GHz\footnote{The actual central frequency of the bands was chosen to be 39.9~GHz, because channels were missing in the first half of each band at the time of our observations. Only one of the observing bands was used in the final data analysis \citep[see also][]{emo11}.}, the frequency that corresponds to the CO(1-0) line ($\nu_{\rm rest}=115.2712$ GHz) Doppler-shifted to $z=1.92$. We observed only above 30$^{\circ}$ elevation and during atmospheric conditions in which there was no significant phase decorrelation of the signal, with T$_{\rm sys} = 50 - 80 {\rm K}$ during the various runs. Short (2 min) scans were made on the nearby strong calibrator PKS~B0130-171 ($S_{\rm 40~GHz} \approx 2.3$ Jy~beam$^{-1}$) roughly every 10 minutes in order to derive frequent gain- and bandpass-calibration solutions, both of which were interpolated in time and applied to the data of MRC~0152-209. We observed Uranus when it was at roughly the same elevation as PKS~B0130-171 in order to flux calibrate our data with an accuracy of 30$\%$. Details on the calibration strategy (using {\sc miriad}) and flux calibration uncertainties are described in \citet{emo11}. 

We used multi-frequency synthesis to obtain a continuum image \citep[with ${\rm robustness} = +1$;][]{bri95}, which we subsequently cleaned. We then separated the spectral-line from the continuum data with a linear fit to the line-free channels in the {\it uv}-domain before Fourier transforming the data into a line-data cube (${\rm robustness} = +1$). The velocity axis was translated to match the optical barycentric rest-frame velocity at the redshift of MRC~0152-209 \citep[$z=1.92$;][]{bre00}. The spatial resolution of our final data is $9.80 \times 7.09$ arcsec (PA\,=\,89.8$^{\circ}$), with a velocity resolution in the line data of 7.6~\kms\ per 1~MHz channel at 39.476~GHz across an effective velocity range of $\sim$15000 \kms. The noise is 58 $\mu$Jy~beam$^{-1}$ in the continuum image and 0.47 mJy~beam$^{-1}$ per 1\,MHz channel in the line data. For the data analysis presented in this paper we have binned the line data to 40 \kms\ wide channels and subsequently applied a hanning smooth, resulting in an effective velocity resolution of 80 \kms\ and a noise level of $\sigma$\,=\,0.16 mJy~beam$^{-1}$\,chan$^{-1}$.

\begin{figure}[t]
\centering
\includegraphics[width=0.45\textwidth]{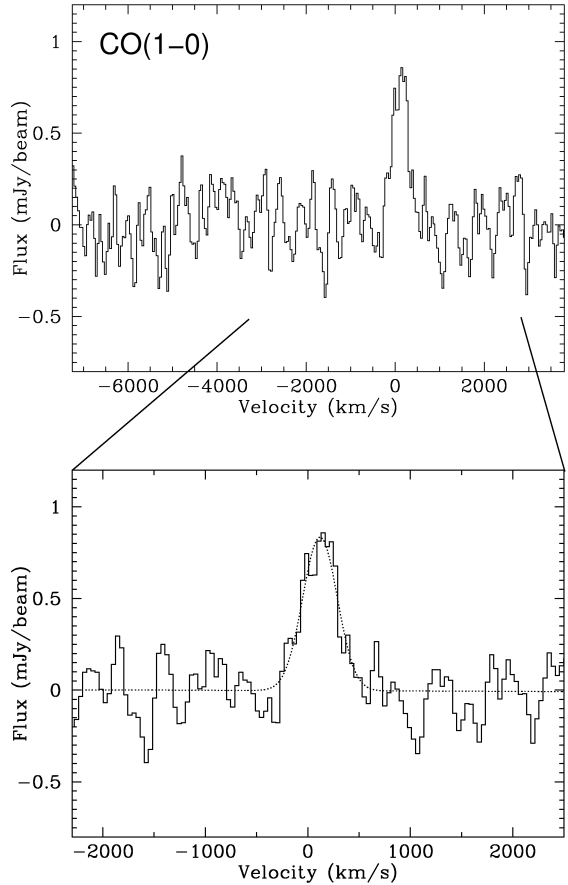}
\caption{{\sl Top:} CO(1-0) spectrum of MRC~0152-209 against the central host galaxy (extracted by averaging the central four pixels of the binned line data, Sect. \ref{sec:observations}). {\sl Bottom:} zoom-in of the CO(1-0) spectrum. The dotted line show a Gaussian fit.}
\label{fig:spectrum}
\end{figure}

\section{Results}
\label{sec:results}

\subsection{CO(1-0)}
\label{sec:resultCO}

\begin{figure*}[htb]
\centering
\includegraphics[width=0.85\textwidth]{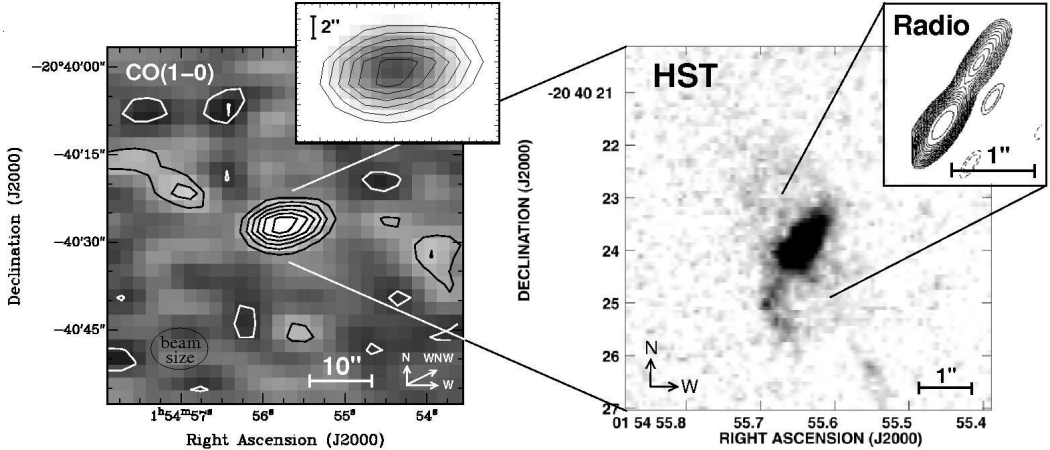}
\caption{{\sl Left:} total intensity image of the CO(1-0) signal of Fig. \ref{fig:spectrum} (see text for details). Contour levels: -3, -2 (white); 2, 3, 4, 5, 6, 7 (black) $\sigma$, corresponding to $L'_{\rm CO}$ = -2.5, -1.7 (white); 1.7, 2.5, 3.3, 4.2, 5.0, 5.8 $\times 10^{10}~{\rm K~km~s^{-1}~pc^2}$. The inset shows the same CO contours overlaid onto the unresolved 40~GHz ATCA radio continuum (grey-scale). {\sl Right:} HST/NICMOS image from \citet{pen01}, showing the disturbed host-galaxy morphology. The inset show the 8.2~GHz radio continuum from \citet{pen00_radio}.}
\label{fig:mom0}
\end{figure*}

Figure \ref{fig:spectrum} shows the spectrum of CO(1-0) against the central region of MRC~0152-209. From fitting a Gaussian to this profile, we derive that the CO(1-0) signal has a peak flux density of $S_{\rm CO} = 0.84 \pm 0.25$ mJy~beam$^{-1}$ \footnote{The error reflects the accuracy of the flux calibration (Sect. \ref{sec:observations})} and ${\rm FWHM} = 397 \pm 37$ \kms. The CO(1-0) signal peaks at a redshift of $z_{\rm CO} = 1.9212 \pm 0.0002$, in agreement with the optical redshift of $z=1.92$ \citep{mcc91,bre00}. Figure \ref{fig:mom0} {\sl (left)} shows a total intensity map of the CO(1-0) line integrated over the channels in which the line was detected. The integrated CO(1-0) signal is detected at the 8$\sigma$ level and $\int_{\rm v} S_{\rm CO}\delta {\rm v} = 0.37 \pm 0.11$ Jy~beam$^{-1}$ $\times$~\kms. There is a tentative indication from Figure \ref{fig:mom0} {\sl (left)} that the CO(1-0) emission is marginally resolved in west-north-western (WNW) direction. This would mean that the CO(1-0) gas is distributed on scales larger than the major beam axis, which corresponds to $\sim$81 kpc. Observations at higher spatial resolution are in progress to verify this.

The CO(1-0) emission-line luminosity can be calculated following \citet[][and references therein]{sol05}:
\begin{equation}
L'_{\rm CO} = 3.25 \times 10^7 (\frac{\int_{\rm v} S_{\rm CO} \delta {\rm v}}{{\rm Jy}~{\rm km/s}}) (\frac{D_{\rm L}}{{\rm Mpc}})^2 (\frac{{\nu_{\rm rest}}}{\rm GHz})^{-2} (1+z)^{-1},
\label{eq:lco}
\end{equation}
with $L'_{\rm CO}$ expressed in ${\rm K~km~s^{-1}~pc^2}$. For MRC~0152-209, $L'_{\rm CO} = 6.6 \pm 2.0 \times 10^{10}\ {\rm K~km~s^{-1}~pc^2}$.

\subsection{Radio continuum}
\label{sec:resultCont}

\begin{figure}[t!]
\centering
\includegraphics[width=0.45\textwidth]{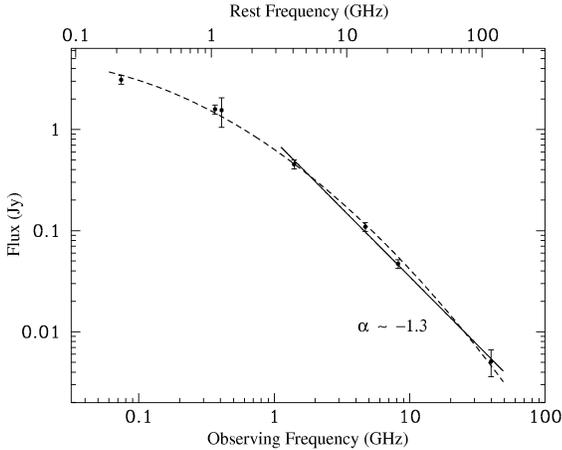}
\caption{Radio spectral index of MRC~0152-209. Data at 1.4, 4.7 and 8.2 GHz are from \citet{pen00_radio}; the lower frequency data are taken from various surveys listed in the NASA Extra-galactic Database (NED -- see http://nedwww.ipac.caltech.edu/). The solid line represents a broken power-law fit ($F_{\nu} \propto {\nu}^{\alpha}$), the dashed line shows a second order polynomial fit.}
\label{fig:mrc0152}
\end{figure}

The radio continuum emission of MRC~0152-209 is unresolved in our ATCA observations and has a flux of $S_{39.9 \rm GHz} = 5.1 \pm 1.5$ mJy. Using our 40\,GHz flux and 8.2, 4.7 and 1.4 GHz data from \citet{pen00_radio}, the spectral index can be fitted with a power-law of the form $S_{\nu} \propto {\nu}^{\alpha}$ from $\sim$1.4 to 40 GHz ($4-115$ GHz in the galaxy's rest-frame; Fig. \ref{fig:mrc0152}). This yields a steep high-frequency spectral index of $\alpha^{\rm 40\,GHz}_{\rm 1.4\,GHz} \approx -1.3$. Data obtained from surveys conducted below 1~GHz suggest a flattening of the spectrum at low frequencies, possibly fitted more accurately by a second order polynomial rather than a single power-law (Fig. \ref{fig:mrc0152}). Additional measurements are required to determine this reliably, but this is beyond the scope of the current paper.

\section{Discussion}
\label{sec:discussion}

\subsection{Molecular gas and star formation}
\label{sec:discussion1}

From the CO(1-0) luminosity we can estimate the mass of molecular gas in MRC~0152-209 by adopting a standard conversion factor $\alpha_{\rm x} = {\rm M}_{\rm H2}/{\rm L}'_{\rm CO}$ [M$_{\odot}$ (K \kms\ pc$^{2}$)$^{-1}$] \citep[where M$_{\rm H2}$ includes a helium fraction; e.g.][]{sol05}. In this paper, we use $\alpha_{\rm x} = 0.8$ M$_{\odot}$ (K \kms\ pc$^{2}$)$^{-1}$ found by \citet{dow98} for ultra-luminous infra-red galaxies (ULIRGs). This also agrees with $\alpha_{\rm x} = 0.8-1.6$ M$_{\odot}$ (K \kms\ pc$^{2}$)$^{-1}$ found for high-$z$ sub-millimeter and star forming galaxies \citep{tac08,sta08}. We note, however, that the conversion of CO luminosity into \Htwo\ mass depends on the properties of the gas, such as metallicity and radiation field \citep[e.g.][]{glo11}, and that conversion factors as high as of $\alpha_{\rm x} = 5$ M$_{\odot}$ (K \kms\ pc$^{2}$)$^{-1}$ have been derived for molecular clouds in the Milky Way \citep[e.g.][]{sol91}. Our assumption of $\alpha_{\rm x} = 0.8$ M$_{\odot}$ (K \kms\ pc$^{2}$)$^{-1}$ thus results in a conservative estimate of the total molecular gas mass in MRC~0152-209 of M$_{\rm H2} = 5 \times 10^{10} M_{\odot}$.

The 24$\mu$m infra-red (IR) flux of MRC~0152-209 ($f_{\rm 24 \mu m} = 3320 \pm 133~\mu$Jy) is among the highest in a large sample of HzRGs observed with {\it Spitzer} by \citet{bre10}. At $z=1.92$, the observed $\lambda=24$ $\mu$m corresponds to $\lambda=8.2$ $\mu$m in the rest-frame, hence from $f_{\rm 24 \mu m}$ we derive a mono-chromatic rest-frame luminosity of ($\nu L_{\nu}$)$^{\rm rest}_{\rm 8.2 \mu m} = 2.7 \times 10^{12} L_{\odot}$. When comparing the mid-IR rest-frame colors \citep{bre10} with IRAC color diagrams by \citet{ste05}, MRC~0152-209 falls within the region that is dominated by active galaxies, indicating that a significant fraction of the 8~$\mu$m flux may originate from a hot dusty torus surrounding the central AGN.

Alternatively, part of ($\nu L_{\nu}$)$^{\rm rest}_{\rm 8.2 \mu m}$ may arise from a dust-enshrouded starburst, which peaks in the far-IR (FIR). \citet{eva05} show that low-$z$ radio galaxies follow a strong correlation between $L'_{\rm CO}$ and $L_{\rm FIR}$. This trend is also seen among high-$z$ submillimeter galaxies \citep[SMGs;][]{gre05,ivi11} and likely reflects the Schmidth-Kennicutt relation between star formation rates and gas reservoir \citep[][]{sch59,ken98}. If this correlation is also valid for HzRGs, the CO(1-0) luminosity of MRC~0152-209 implies a FIR luminosity of order $L_{\rm FIR} \sim 5 \times 10^{12}$ $L_{\odot}$ \citep{ivi11}, well in the regime of ULIRGs. \citet{bav08} show that the mono-chromatic IR luminosities of starburst-dominated ULIRGs at 8, 24, 70 and 160$\mu$m are strongly correlated with the total IR luminosity ($L_{\rm IR} = L_{5-1000 \mu m}$), with reliable correlations up to $z \sim 2$. While these correlations may not necessarily be valid in case a significant fraction of ($\nu L_{\nu}$)$^{\rm rest}_{\rm 8.2 \mu m}$ in MRC~0152-209 is AGN related, we argue that they still provide a reliable upper limit to $L_{\rm IR}$, given that a `starburst-only' scenario will provide the highest possible energy output in the FIR and thus maximize $L_{\rm IR}$. From ($\nu L_{\nu}$)$^{\rm rest}_{\rm 8 \mu m}$ and following \citet[][their Equation 6]{bav08}, we therefore estimate that $L_{\rm IR} \lesssim 7.9 \times 10^{12} L_{\odot}$ for MRC~0152-209. 

The upper limit on the dust-to-gas ratio in MRC~0152-209 can then be estimated to be $L_{\rm IR}/L'_{\rm CO} \lesssim 120$, consistent with that of nearby ULIRGs \citep{sol97}. Following \citet{ken98}, the upper limit on the star formation rate is ${\rm SFR} \approx \frac{L_{\rm IR}}{5.8 \times 10^{9} L_{\odot}} \lesssim 1362~{\rm M}_{\odot}$~yr$^{-1}$. Assuming that the molecular gas in MRC~0152-209 offers a reservoir to sustain this maximum star formation rate, the lower limit on the gas depletion time-scale is t$_{\rm depletion} = \frac{{\rm M}_{\rm H2}}{\rm SRF} \gtrsim \epsilon^{-1} \cdot 39$ Myr (with $0 \leq \epsilon \leq 1$ the star formation efficiency). According to models by \citet{mih94}, this minimum gas depletion time-scale is comparable to the typical lifetime of a massive burst of star formation during a major merger, which could give the merging system the appearance of a ULIRG for about $\sim 50$ Myr (although \citet{swi06} show that for high-$z$ SMGs a starburst age of several 100 Myr is more realistic). It is conceivable that also MRC~0152-209 has undergone a recent merger, given its disturbed optical morphology \citep{pen01}.

Therefore, MRC~0152-209 shares characteristics commonly found among ULIRGs, but its classification remains somewhat ambiguous until the nature of the IR emission and star formation properties are better understood. It is clear, however, that MRC~0152-209 contains a large reservoir of molecular gas that has not (yet) been depleted by vigorous star formation.

\subsection{CO(1-0) in HzRGs}
\label{sec:discussion2}

The CO(1-0) luminosity of MRC~0152-209 is an order of magnitude larger than that found in nearby radio galaxies \citep{oca10,eva05}, indicating that MRC~0152-209 contains significantly more molecular gas than its low-$z$ counterparts. 

Assuming $\alpha_{\rm x} = 0.8$, the H$_{2}$ mass of MRC~0152-209 is of the same order as that derived from CO detections in other HzRGs \citep[][]{sco97,pap00,pap01,all00,bre03,bre03AR,bre05,gre04,kla05,ivi08,nes09} as well as from a survey of high-$z$ SMGs \citep[][]{gre05}. However, most of these CO studies relied on observations of the higher rotational CO transitions, which could underestimate the total mass and spatial extent of the molecular gas (see Sect. \ref{sec:introduction}). 

To our knowledge, CO(1-0) detections have been claimed for only two other HzRGs, namely TN~J0924-2201 \citep[$z=5.2$;][]{kla05}
and 4C~60.07 \citep[$z=3.8$;][]{gre04,ivi08}. Both systems contain M$_{\rm H2} \sim 10^{11} M_{\odot}$. Interestingly, they both are IR-faint objects \citep{bre10,ivi08}. CO(1-0) observations of four SMGs \citep[$z=2.2-2.5$;][]{ivi11} and three massive star-forming disk ($BzK$)  galaxies \citep[$z\sim1.5$;][]{ara10} reveal CO(1-0) luminosities comparable to that of MRC~0152-209. {\sl CO(1-0) observations of larger and unbiased samples of various types of high-$z$ galaxies are necessary to investigate and compare the molecular gas content in these systems in a systematic way.}

\subsubsection{CO(1-0) and radio source properties}
\label{sec:discussion3}

The radio source in MRC~0152-209 is fairly compact (13 kpc) and not reaching beyond the optical boundaries of the host galaxy \citep{pen00_radio,pen01}. Based on a CO(3-2) detection in the outer part of the \lya\ halo that surrounds the $z = 2.6$ radio galaxy TXS~0828+193, \citet{nes09} speculate that large amounts of dense gas (and perhaps dust) may be present in the halos of HzRGs at radii that are not yet affected by mechanical heating from the radio source. This would be in agreement with studies of the neutral gas in HzRGs by \citet{oji97_HI}, who find that strong \HI\ absorption is predominantly associated with small ($<50$kpc) radio sources. It will be particularly interesting to investigate whether the CO(1-0) gas, i.e., the part of the observable molecular gas that is least affected by excitation and heating, shows a similar trend with radio source size, or whether {\sl on average} HzRGs are fainter in CO(1-0) than their radio-quiet high-$z$ counterparts.

Besides MRC~0152-209, we recently published a tentative CO(1-0) detection with ATCA/CABB in the HzRG MRC~0943-242 \citep{emo11}. This tentative CO(1-0) detection is found in the outer part of the \lya\ halo that surrounds MRC~0943-242 (i.e. beyond the $\sim$15~kpc radius of the radio source). Regarding the other two known HzRGs with CO(1-0) detections (Sect. \ref{sec:discussion2}), TN~J0924-2201 also contains a compact radio source \citep[][]{kla05}, while for 4C~60.07 the radio source is extended, but the CO(1-0) most likely represents tidal debris from a galaxy-galaxy interaction \citep[$z=3.8$;][]{ivi08}.

In a forthcoming paper, we will present result of our ongoing systematic CO(1-0) survey of a larger sample of HzRG with ATCA/CABB. This will allow further investigation into possible connections between radio-AGN activity and the cold molecular gas properties of HzRGs.

\section{Conclusions}
\label{sec:conclusions}

We have presented a robust detection of CO(1-0) in the high-$z$ radio galaxy MRC~0152-209 ($z=1.92$), tracing a molecular gas mass of M$_{\rm H2} = 5 \times 10^{10}$ ($\alpha_{\rm x}$/0.8) M$_{\odot}$. MRC~0152-209 is an IR-bright radio galaxy that contains a large reservoir of cold molecular gas that has not (yet) been depleted by star formation or radio source feedback. 

CO(1-0) is the most robust tracer of molecular gas (including the wide-spread, low-density and sub-thermally excited component) in high-$z$ galaxies. The search for CO(1-0) in HzRGs at observing frequencies below 50~GHz with the ATCA and EVLA will be a crucial complement to surveys of the higher CO transitions with ALMA.

\acknowledgments
We thank Laura Pentericci for the use of her HST/NICMOS and VLA radio continuum images. Reproduced by permission of the AAS. The Australia Telescope is funded by the Commonwealth of Australia for operation as a National Facility managed by CSIRO. This research has made use of the NASA/IPAC Extragalactic Database (NED) which is operated by the Jet Propulsion Laboratory, California Institute of Technology, under contract with the National Aeronautics and Space Administration.


\end{document}